\documentclass[showkeys,aip,graphicx,amsmath,superscriptaddress,amssymb]{revtex4-2}
\usepackage{graphicx}
\usepackage{hyperref}
\usepackage[capitalize]{cleveref}
\usepackage{xcolor}
\usepackage{array}

\begin{document}
\title{Exascale-ready HemeLB for large-eddy simulation of full-human arterial blood flow}
%
%\titlerunning{Accurate full-human HemeLB with LES}
\author{Xiao Xue}
\affiliation{ 
Centre for Computational Science, University College London, London, UK
}
\author{Ioannis Zacharoudiou}%
\affiliation{ 
Centre for Computational Science, University College London, London, UK
}
\author{Peter V. Coveney}%
\email{p.v.coveney@ucl.ac.uk}
\affiliation{ 
Centre for Computational Science, University College London, London, UK
}
\affiliation{
Centre for Advanced Research Computing, University College London, UK
}

\date{\today}

\begin{abstract}
Direct numerical simulation (DNS) of blood flow through the whole human arterial tree is limited by the number of lattice sites that must be advanced on a high-performance computer. We address this limit by combining the lattice Boltzmann solver HemeLB with Smagorinsky large-eddy simulation (LES). HemeLB runs with the same scaling behaviour on AMD, Intel and NVIDIA graphics processing units (GPUs) and sustains about 10 trillion lattice-site updates per second (TLUPS) on both the Frontier and Aurora exascale supercomputers. In stenotic aortic flow, LES on lattices five and ten times coarser than a 20~$\mu$m DNS reproduces the wall pressure and the cross-sectional pressure and velocity, and its plane-averaged pressure differs from that of the DNS by at most 0.75~mmHg, below 1\% of the local pressure. At 100~$\mu$m, LES reduces the estimated cost of a full-human simulation by a factor of about 625 relative to DNS. We use this resolution to compute blood flow from the head to the feet in a full-human arterial model with 295 million fluid sites, one inlet at the aortic root and 432 Windkessel outlets, at about 15--18 node-hours per cardiac cycle on two Frontier nodes. A single simulation resolves every region simultaneously, from the cerebral arteries to the arteries of the feet. The pressure pulse grows as it travels from the aorta to the feet, the shear stress concentrates in the branches rather than in the aorta, the pulse reaches the outlets later with increasing distance from the heart, and the flow in the legs and feet is triphasic, as in Doppler measurements of the human leg. These results bring patient-specific whole-body lattice Boltzmann simulations within reach of current supercomputers and pave the way for ensemble simulations towards a standard human model.

\end{abstract}

\maketitle    
\section{Introduction}
% motivation
Cardiovascular disease (CVD) is the leading cause of death globally~\cite{who_cvd_2020}. Pressure, velocity and wall shear stress in the arteries help to explain how CVD develops and how it responds to treatment~\cite{morris2016computational,niederer2019computational}. Abnormal wall shear stress, for example, is associated with the initiation, growth and rupture of aneurysms~\cite{meng2014high}. Bedside hemodynamic monitoring, however, provides global or point measurements such as cardiac output rather than spatially resolved flow fields~\cite{huygh2016hemodynamic}. Image-based computational fluid dynamics (CFD) addresses this limitation by reconstructing patient-specific vessel geometry from computed tomography or magnetic resonance images and computing the flow non-invasively~\cite{antiga2008image,morris2016computational,caballero2013review,quarteroni2019mathematical}. It complements four-dimensional flow magnetic resonance imaging, which measures blood velocity in vivo~\cite{kamada2022blood}. CFD predictions depend on the quality of the reconstructed geometry~\cite{geers2011patient} and on the imposed inlet and outlet boundary conditions~\cite{marzo2011computational,madhavan2018effect,youssefi2018impact}. Most such studies model a single arterial segment and represent the rest of the circulation through boundary conditions, for example when assessing stent implantation~\cite{morris2016computational,chiastra2016computational} or aortic dissection~\cite{bonfanti2019patient}. A connected model of the arterial network, extending to the hands and feet, would allow local flow features to be studied in their systemic context. Such models represent a major step towards virtual humans, or human digital twins~\cite{hoekstra2018virtual,coveney2023virtual,lo2025multi}.

% full-body arterial modelling
Whole-body arterial hemodynamics has long been studied with one-dimensional network models, which describe pressure and flow wave propagation along each vessel~\cite{shi2011review,formaggia2006numerical}. Xiao et al.~\cite{xiao2013multi} presented a three-dimensional model of hemodynamics in a deformable network of the major arteries from head to feet without resolving the distal vessels. Its 82 outlets were coupled to three-element Windkessel models, and each cardiac cycle took about 48~h on 384 central processing unit (CPU) cores~\cite{xiao2013multi}. Wall deformability can be captured through fluid--structure interaction~\cite{bazilevs2009patient}, and the impact of this and other modelling assumptions on aortic hemodynamics has been assessed~\cite{alastruey2016impact}. Beyond the outlets of such three-dimensional domains, the distal circulation is commonly represented by lumped-parameter Windkessel models of downstream resistance and compliance~\cite{westerhof2009arterial,capoccia2015development,grinberg2008outflow}. Such three-dimensional--zero-dimensional coupling has also been used with cardiac models~\cite{augustin2021computationally} and with lattice Boltzmann simulations of aortic flow~\cite{wei2023direct}. These studies show that full-body three-dimensional simulation is feasible, but its cost constrains the spatial resolution, the extent of the vascular tree and the simulated time.

% exascale hardware and the lattice Boltzmann method
Reducing this cost depends on both the hardware and the numerical method. Exascale supercomputers such as Frontier deliver more than a billion billion ($10^{18}$) floating-point operations per second, most of it from GPUs~\cite{atchley2023frontier}. Using them efficiently requires algorithms with local data access and little communication. Most cardiovascular CFD solves the incompressible Navier--Stokes equations~\cite{morris2016computational,quarteroni2019mathematical}. These solvers typically require a global pressure solution at every time step, which makes parallel scaling demanding. Finite-element multiphysics codes have nevertheless been scaled to about 100,000 CPU cores~\cite{vazquez2016alya}. The lattice Boltzmann method (LBM) instead evolves particle distribution functions whose moments recover the Navier--Stokes equations at low Mach number~\cite{succi2001lattice,Kruger2017}. Its collision step is local and its streaming step exchanges data only with nearest neighbours, so it maps efficiently onto large GPU systems~\cite{Calore2016,rinaldi2012lattice}. For arterial flow, LBM has been compared directly with the finite-element method~\cite{axner2009simulations}. LBM has also been applied to flows ranging from high-Reynolds-number turbulence~\cite{hou1994lattice,shao2022near} to colloidal particle dynamics~\cite{ladd1993short,xue2020brownian,xue2021lattice}.

% HemeLB and the cost of full-human resolution
HemeLB is an open-source LBM solver designed for the sparse geometries of vascular networks~\cite{mazzeo2008hemelb,hemeLB}, and it has been verified and validated for vascular flows~\cite{nash2014choice,groen2018validation,mccullough2021high}. It has been used to study the circle of Willis~\cite{mccullough2023high}, arteriovenous fistulae~\cite{mccullough2021high}, abdominal aortic aneurysms~\cite{lo2024uncertainty,xue2026uncertainty}, stenotic aortic flow~\cite{xue2024lattice} and aortic flow coupled to cardiac electromechanics~\cite{lo2025multi}. Other LBM studies have addressed flow in aneurysmal arteries~\cite{afrouzi2020simulation} and in coarctation of the aorta~\cite{feiger2020accelerating}. At the outlets, HemeLB imposes pressures computed from Windkessel models~\cite{nash2014choice,lo2025multi}, and an efficient boundary condition has been developed to control outlet flow rates in large arterial networks~\cite{lo2022parametric}. The GPU version of HemeLB was developed for large scale blood flow simulation~\cite{zacharoudiou2023development}. Its efficiency on an exascale system such as Frontier, with four AMD Instinct MI250X GPUs per node~\cite{atchley2023frontier}, has not yet been reported. Exascale capability alone, however, does not make full-human simulation useful in practice. Because HemeLB voxelises the geometry onto a uniform lattice~\cite{lo2025multi}, one spacing must satisfy two requirements over the whole domain. It must place enough sites across the narrowest retained vessels. Without a turbulence model, it must also resolve the smallest flow scales in the aorta, where the Reynolds number can exceed 4600~\cite{xue2024lattice}. The number of fluid sites scales with the inverse cube of the lattice spacing. The time step must also shrink in proportion to the spacing to keep the Mach number small, so the cost per simulated second scales with the inverse fourth power of the spacing. A thoracic aorta alone needs about 9 billion sites at 20~$\mu$m~\cite{xue2024lattice,lo2025multi}. A lattice of this size occupies thousands of nodes of an exascale machine, and a full-human model is larger still. Patient-specific studies need many such runs. They must simulate several cardiac cycles to reach periodicity~\cite{xiao2013multi}, calibrate outlet models~\cite{lo2022parametric} and sample parameter uncertainty with ensembles~\cite{lo2024uncertainty}. The cost per simulation must therefore fall before exascale-ready HemeLB becomes useful for patient-specific full-human studies.

% LES and the present approach
LES offers this reduction by relaxing the second requirement. It resolves the large flow structures and represents the unresolved scales with a subgrid-scale (SGS) model~\cite{Sagaut2002}. Coarsening the lattice without such a model is not an option in LBM. On a coarser lattice the dimensionless lattice viscosity is smaller, the relaxation time $\tau$ approaches its stability limit of $1/2$, and the simulation becomes unstable once the flow contains scales that the lattice cannot resolve~\cite{Kruger2017}. The SGS model adds an eddy viscosity where the resolved strain rate is large, which removes energy from the unresolved scales and keeps the scheme stable~\cite{hou1994lattice}. For wall-bounded turbulence, the number of grid points required by LES grows more slowly with Reynolds number than that required by DNS~\cite{choi2012grid,yang2021grid}. LES is well established for turbulent engineering flows such as wind-farm aerodynamics~\cite{porte2011large,mehta2014large}. In the arterial tree, helical and retrograde flow arises in the aortic arch~\cite{kilner1993helical,leuprecht2003blood,caballero2013review}, and transitional flow can develop distal to stenoses~\cite{ding2021transitional,xue2024lattice}. LES does not relax the first requirement, so it reduces cost only where the flow, rather than the geometry, sets the spacing. In a full-human model, the narrowest retained vessels permit a spacing several times coarser than that needed to resolve the aortic flow directly. LES allows the whole model to run at this coarser spacing. Because both the number of fluid sites and the number of time steps fall as the spacing grows, the cost then falls by orders of magnitude, provided that LES retains the accuracy of the resolved hemodynamics.

% validation case and its scope
We test this proviso in stenotic aortic flow. The stenosis accelerates the flow and can make it transitional distal to the narrowing~\cite{ding2021transitional,xue2024lattice}, which is where a subgrid model is most likely to misrepresent the flow. Agreement with DNS on this geometry is therefore a stringent test of the LES in the aorta. We also test a lattice coarser than the narrowest vessels of a full-human model permit, to establish how far the aortic flow alone tolerates coarsening. This validation concerns the aorta, where the flow sets the spacing. In the narrowest vessels of the full-human model, which span only a few lattice spacings, the accuracy is limited by the geometric resolution rather than by the subgrid model, and it is not addressed by this validation.

% choice of subgrid model and the present approach
Within LBM, LES has been formulated with Smagorinsky~\cite{hou1994lattice,Eggels1996a,Ansumali2004}, dynamic~\cite{Premnath2009} and approximate-deconvolution~\cite{Malaspinas2011} subgrid models. The dynamic and deconvolution models require explicit filtering over neighbouring lattice sites~\cite{Premnath2009,Malaspinas2011}. In the Smagorinsky model~\cite{smagorinsky1963general}, the strain rate is instead computed locally from the non-equilibrium part of the distribution functions~\cite{hou1994lattice}. It therefore adds no communication cost and preserves the locality that makes LBM efficient on GPUs. Here we combine the GPU implementation of HemeLB with Smagorinsky LES and Windkessel outlet models~\cite{lo2025multi}. We ask how far LES can reduce the cost of a full-human simulation on exascale GPU systems without sacrificing accuracy, and thus bring such simulations closer to patient-specific use.

\begin{figure}
\includegraphics[width=\textwidth]{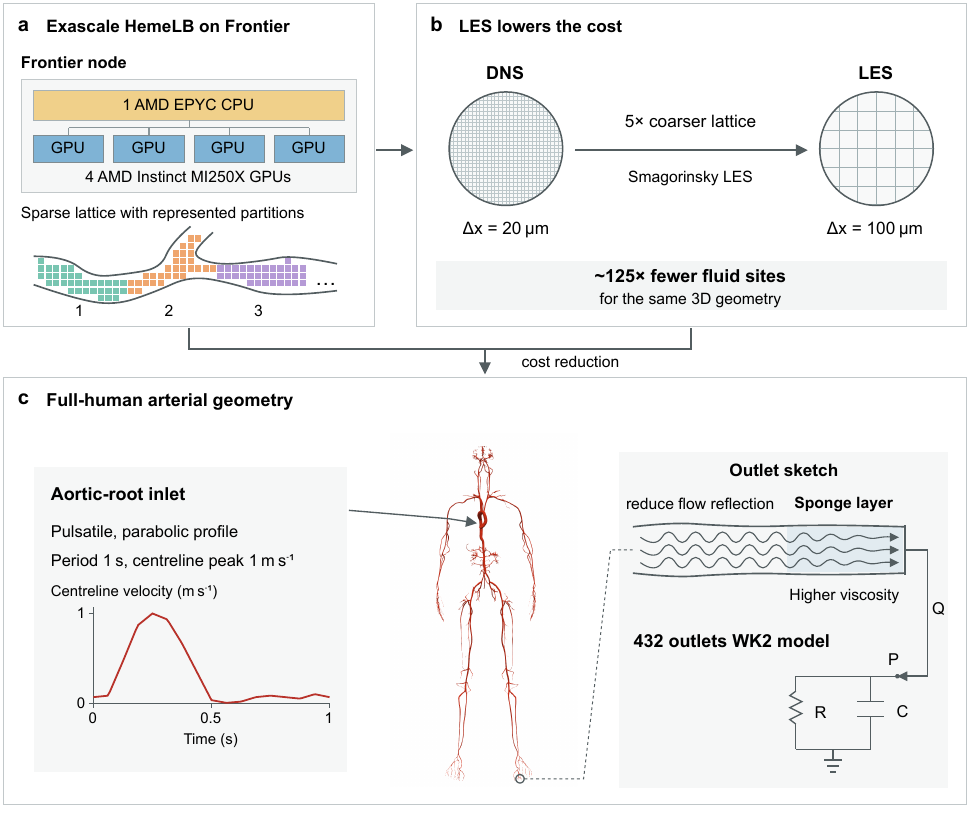}
\caption{\textbf{Exascale HemeLB and LES for full-human arterial blood flow.}
\textbf{a}, A Frontier node has one AMD EPYC CPU and four AMD Instinct MI250X GPUs. HemeLB divides the sparse vascular lattice into many partitions, which are processed in parallel across the GPUs.
\textbf{b}, Smagorinsky LES allows a lattice five times coarser than DNS, from $\Delta x=20$~$\mu$m to $\Delta x=100$~$\mu$m, which reduces the number of fluid sites by about 125 times for the same three-dimensional geometry.
\textbf{c}, Full-human arterial geometry. Blood enters at the aortic root with a pulsatile velocity of parabolic profile and leaves through 432 outlets. The inset shows the imposed centreline velocity over one cardiac cycle of 1~s, with its peak of 1~m\,s$^{-1}$ at $t=0.25$~s. The outlet sketch shows that, before each outlet, a sponge layer of higher viscosity damps velocity fluctuations and reduces flow reflections. The outlet flow rate $Q$ then sets the outlet pressure $P$ through a two-element Windkessel (WK2) model with resistance $R$ and compliance $C$.}
\label{figure:wholehuman}
\end{figure}

% contributions
To answer this question, we measure the throughput, strong scaling and parallel efficiency of HemeLB on the Frontier and Aurora exascale systems~\cite{atchley2023frontier,allcock2025aurora} and on the NVIDIA-based Summit and Polaris systems~\cite{vazhkudai2018design,alcf2026polaris}. These measurements convert a lattice size into node-hours, and so give the cost against which the accuracy of LES is weighed. Unlike its first GPU version~\cite{zacharoudiou2023development}, which ran only on NVIDIA GPUs, HemeLB now runs on NVIDIA, AMD and Intel GPUs from a single code base. We then quantify the accuracy--cost trade-off of Smagorinsky LES in stenotic aortic flow, where pressure and velocity on lattices five and ten times coarser than a 20~$\mu$m DNS are compared with the DNS and related to fluid-site count and computational cost. With the validated LES, we compute blood flow through a full-human three-dimensional arterial tree from the head to the feet. This single simulation resolves every region at once and shows that the pressure pulse grows from the aorta to the feet, that the wall shear stress concentrates in the branches rather than in the aorta, that the pulse reaches the outlets later with increasing distance from the heart, and that the flow in the legs and feet is triphasic. \Cref{figure:wholehuman} summarises this approach.

The remainder of this paper is organised as follows. \Cref{sec:methods} describes the numerical methods. \Cref{sec:performance} describes the platform-portable implementation and reports the performance of HemeLB on AMD, Intel and NVIDIA GPU systems, including the Frontier and Aurora exascale systems. \Cref{sec:aorta-validation} validates the LES on stenotic aortic flow. \Cref{sec:fullhuman} presents the full-human LES of the arterial system and discusses the limitations of the present model. \Cref{sec:conclusion} summarises the main findings. The symbols and abbreviations used throughout the paper are listed in \cref{app:notation}.

\section{Methods}\label{sec:methods}

\subsection{The lattice Boltzmann method}\label{sec:method-lbm}

%Note that the momentum flux, $\mathbf{\Pi}(\mathbf{x}, t)$, can be presented by the sum of the equilibrium and non-equilibrium parts, $\mathbf{\Pi}(\mathbf{x}, t)  = \mathbf{\Pi}_{\mathrm{eq}} (\mathbf{x}, t)  + \mathbf{\Pi}_{\mathrm{neq}}(\mathbf{x}, t)$. 
This study employs a three-dimensional lattice Boltzmann model featuring 19 discretised directions, known as the D3Q19 model~\cite{Kruger2017}. The fluid domain is discretised on a uniform cubic lattice with spacing $\Delta x$. Each lattice cell is specified by its position $\mathbf{x}$ and time $t$, and is characterized by a discretised velocity set $\mathbf{c}_i$ where $i \in \{0, 1, \ldots, 18\}$. The evolution equation for the distribution functions can be written as:
\begin{equation}
\label{eq:lbe}
f_i(\mathbf{x}+\mathbf{c}_{i}\Delta t,t+\Delta t) =f_i(\mathbf{x}, t) - \Omega \left[f_i(\mathbf{x},t )-f_i^{eq}(\mathbf{x},t )\right],
\end{equation}
where $\Omega$ denotes the  Bhatnagar--Gross--Krook (BGK), or single-relaxation-time, collision kernel~\cite{Bhatnagar1954,succi2001lattice}, defined as 
$\Omega = \frac{\Delta t}{\tau}$. $\tau$ is the relaxation time. The collision kernel relaxes the distribution function towards the local Maxwellian distribution function $f_i^{eq}$:
\begin{equation}
\label{eq:local_eq}
f_i^{eq}\left(\mathbf{x},t\right) =  w_{i}\rho\left(\mathbf{x},t\right) \bigg[1+\frac{\mathbf{c}_i\cdot\mathbf{u}\left(\mathbf{x},t\right)}{c_s^2}+\frac{\left[\mathbf{c}_i\cdot\mathbf{u}\left(\mathbf{x},t\right)\right]^2}{2c_s^4}  - \frac{\left[\mathbf{u}\left(\mathbf{x},t\right)\cdot\mathbf{u}\left(\mathbf{x},t\right)\right]}{2c_s^2}\bigg] \;,
\end{equation}
where $w_{i}$ denotes the direction based weights ($w_{i}$ = 1/3 for i=0, 1/18 for the six nearest neighbours and 1/36 for the remaining directions), $\rho\left(\mathbf{x},t\right)$ is the cell macroscopic density, $\mathbf{u}\left(\mathbf{x},t\right)$ is the cell macroscopic velocity. In~\cref{eq:lbe}, $\Delta t$ symbolizes the LBM time step, which is set to unity. The LBM kinematic viscosity $\nu$ is defined as

\begin{equation}
\label{eq:nu}
\nu = c_s^2\left(\tau -\frac{1}{2}\right)\Delta t,
\end{equation}
with $c_s$ representing the speed of sound, and $c_s^2$ equating to $1/3$ in lattice units (LU). 
 Macro-scale quantities such as density and momentum are derived from the moments of the distribution function $f_i(\mathbf{x}, t)$, the discrete velocities $\mathbf{c}_{i}$:
\begin{equation}\label{eq:density}
\rho(\mathbf{x}, t) = \sum_{i=0}^{18} f_i(\mathbf{x}, t), \\
\end{equation} 
\begin{equation}
\label{eq:momentum}
\rho(\mathbf{x}, t)\mathbf{u}(\mathbf{x}, t) = \sum_{i=0}^{18} f_i(\mathbf{x}, t)\mathbf{c}_{i},
\end{equation}

\subsection{Smagorinsky subgrid-scale modelling}\label{sec:method-sgs}
We briefly summarize the LBM-based Smagorinsky SGS LES technique: Within the LBM framework, the effective viscosity $\nu _{\mathrm{eff}}$ \cite{smagorinsky1963general,hou1994lattice} is modelled as the sum of the molecular viscosity, $\nu_0$, and the turbulent viscosity, $\nu_t$:

\begin{equation}
\label{eq:sgs}
\nu _{\mathrm{eff}}=\nu _0+\nu_t, \hspace{.6in} \nu_t = C_{\mathrm{smag}}\Delta ^2\left |\bar{\mathbf{S}}\right |,
%\label{smogrinsky_model}
\end{equation}
where $\left |\mathbf{\bar{S}} \right |$ is the magnitude of the filtered strain rate tensor, $C_{\mathrm{smag}}$ is the Smagorinsky constant, $\Delta$ represents the filter size. We apply~\cref{eq:sgs} into~\cref{eq:nu} to add the turbulent viscosity. The detailed demonstration of the turbulent viscosity can be found in ~\cite{hou1994lattice,xue2022synthetic}.

\subsection{Windkessel outlet boundary conditions}\label{sec:method-wk}
Each outlet of the full-human model is coupled to a WK2 model~\cite{westerhof2009arterial,shi2011review}, also known as a resistance--capacitance outflow condition~\cite{grinberg2008outflow}, as sketched in \cref{figure:wholehuman}. The WK2 model represents the vasculature downstream of the outlet by a peripheral resistance $R$ in parallel with a compliance $C$~\cite{westerhof2009arterial}. The outlet pressure $P$ and the outward volumetric flow rate $Q$ obey
\begin{equation}
\label{eq:wk2}
C\frac{\mathrm{d}P}{\mathrm{d}t}+\frac{P}{R}=Q,
\end{equation}
where $P$ is the gauge pressure relative to the reference state of the lattice. We take this reference pressure as 90~mmHg, so the pressures of the full-human simulation are reported as $P+90$~mmHg. \Cref{eq:wk2} is discretised with the implicit (backward) Euler scheme,
\begin{equation}
\label{eq:wk2_discrete}
P^{n+1}=\frac{RC}{RC+\Delta t}P^{n}+\frac{R\,\Delta t}{RC+\Delta t}Q^{n+1},
\end{equation}
where $\Delta t$ is the lattice Boltzmann time step expressed in physical units. After each time step, $Q^{n+1}$ is computed by summing the outward normal velocity over the lattice sites adjacent to the outlet, and $P^{n+1}$ is imposed during the following step. The pressure is converted to a lattice density, $\rho_{\mathrm{out}}=1+P/c_s^2$ in LU, and the distribution functions that enter the fluid through the outlet are replaced by their equilibria (\cref{eq:local_eq}) at $\rho_{\mathrm{out}}$, as in the pressure boundary condition of Nash et al.~\cite{nash2014choice}.

\subsection{Sponge layer implementation}\label{sec:sponge-sgs}
%TODO:
 To damp velocity fluctuations near each outlet and reduce flow reflection from the outlet boundary, a {\em sponge zone} is set up as a damping region before the outlet~\cite{guo1994comparison,adams1998direct,vergnault2012lattice,xue2022synthetic,xue2023wall}, as sketched in \cref{figure:wholehuman}. In the sponge zone, we impose a higher kinematic viscosity, $\nu_{s}$, defined as
\begin{equation}
\label{eq:sponge_visco}
% vRatioTot *= 1.0 + (vRatio - 1.0) * (dist / width - 1.0) * (dist / width - 1.0);
\nu_{s} = \nu_{\mathrm{eff}}\left [1 + (p - 1) \left (1-\frac{\left \|d  \right \|}{w_{\mathrm{width}}}\right)^2\right],\qquad \left \|d  \right \|\le w_{\mathrm{width}},
\end{equation}
where $p$ is an empirical viscosity ratio, set to $p=1000$ in this work as in previous LBM blood-flow studies~\cite{xue2024lattice,lo2025multi}. $\left \|d  \right \|$ denotes the distance of the lattice cell point to the outlet plane. $w_{\mathrm{width}}$ represents the width of the sponge zone prior to the outlet. The viscosity therefore rises from $\nu_{\mathrm{eff}}$ at the start of the sponge zone to $p\,\nu_{\mathrm{eff}}$ at the outlet.

\section{Computational performance on exascale hardware}\label{sec:performance}
This section describes the changes that make HemeLB run on AMD, Intel and NVIDIA GPUs and measures its strong scaling on these GPUs, including the Frontier and Aurora exascale supercomputers, where it sustains about 10~TLUPS. It then uses this throughput to estimate the cost of a full-human simulation and shows that LES at 100~$\mu$m reduces this cost by a factor of about 625 relative to DNS at 20~$\mu$m.

% \textcolor{blue}{Ioannis: please add 1-2 paragraphs or with an algorithm table for this section how do we merge with different platform}
\subsection{Platform-portable implementation}\label{sec:portability}
% \XX{[Ioannis: Code changes relative to the published GPU version:}
The first GPU version of HemeLB~\cite{zacharoudiou2023development} was written in CUDA C++ and therefore ran only on NVIDIA GPUs, although that work already reported that a port to HIP for AMD GPUs was under way. The present version is platform-agnostic and runs on NVIDIA, AMD and Intel GPUs. We factored the CUDA-specific parts of the code out into back-ends for CUDA (NVIDIA), HIP (AMD) and SYCL (Intel), while the rest of the code is kept in a single branch with one set of source files. This design reduces code duplication and keeps the maintenance effort low across the three platforms. Further changes accompany the move to portability and improve both the performance and the functionality of the code. As in the first version, blood flow is driven by imposing pressure or velocity boundary conditions at the inlets and outlets of the domain. The velocity boundary conditions have been improved and now match the computational performance of the pressure boundary conditions. The geometry input file is now read with MPI I/O, which reduces the initialisation time of a simulation. Finally, functionality previously available only in the CPU version has been ported to the GPU version, including checkpointing, which makes long simulations feasible.

\begin{figure}
\includegraphics[width=\textwidth]{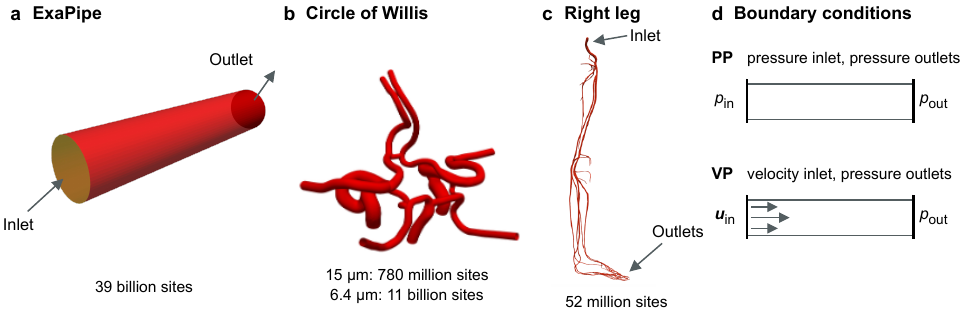}
\caption{\textbf{Benchmark geometries and boundary-condition set-ups.} (\textbf{a}) ExaPipe benchmark, a straight pipe with one inlet and one outlet; the render is adapted from Lo et al.~\cite{lo2025multi}. (\textbf{b}) Circle of Willis. (\textbf{c}) Right leg, with one inlet and outlets at the vessel ends. The number of lattice sites is given below each geometry. (\textbf{d}) PP and VP set-ups, sketched on a straight vessel segment: pressure $p_{\mathrm{in}}$ or velocity profile $\boldsymbol{u}_{\mathrm{in}}$ at the inlet, and pressure $p_{\mathrm{out}}$ at the outlets. Renders are not to scale.}
\label{figure:benchmarks}
\end{figure}
\subsection{Platforms and performance measurement}\label{sec:hpc}
Parallel performance is measured with the GPU implementation of HemeLB~\cite{zacharoudiou2023development} on two systems at the Oak Ridge Leadership Computing Facility. Frontier is an exascale supercomputer~\cite{atchley2023frontier}, and Crusher is its early-access test system with the same node architecture. Each node of both systems has one AMD EPYC CPU and four AMD Instinct MI250X GPUs. Performance is reported as the aggregate number of lattice-site updates per second over all nodes, in million lattice-site updates per second (MLUPS). For the largest runs it is given in TLUPS, where 1~TLUPS equals 1 million MLUPS. Each data series keeps the problem size fixed while the number of nodes increases, which measures strong scaling. \Cref{figure:benchmarks} shows the three benchmark geometries. The ExaPipe benchmark is a straight cylindrical pipe with one inlet and one outlet, and it has 39 billion lattice sites. This size was chosen to be close to the 37 billion fluid sites of the full-human arterial tree at 20~$\mu$m, so ExaPipe stands in for a full-human DNS. The circle of Willis~\cite{mccullough2023high} is a sparse cerebral arterial network, with 780 million sites at 15~$\mu$m and 11 billion sites at 6.4~$\mu$m. The right leg is the arterial tree of a human lower limb, with one inlet and outlets at the vessel ends, and it has 52 million sites. On Crusher, the test problems are the right leg and the circle of Willis at 15~$\mu$m. On Frontier, they are the circle of Willis at 6.4~$\mu$m and ExaPipe. Measured performance is compared with ideal linear scaling and with 75\% of ideal. To compare GPU architectures, the circle of Willis at 6.4~$\mu$m and the ExaPipe benchmark are also run on Summit~\cite{vazhkudai2018design} at the Oak Ridge Leadership Computing Facility, with six NVIDIA V100 GPUs per node, and on Polaris~\cite{alcf2026polaris} and Aurora~\cite{allcock2025aurora} at the Argonne Leadership Computing Facility, with four NVIDIA A100 GPUs and six Intel Data Center GPU Max GPUs per node, respectively. Two boundary-condition set-ups are compared, and \cref{figure:benchmarks} sketches both. In the pressure--pressure (PP) set-up, pressures are prescribed at the inlet and at the outlets. In the velocity--pressure (VP) set-up, a velocity profile is prescribed at the inlet and pressures at the outlets. The circle of Willis at 6.4~$\mu$m is run with both set-ups. The circle of Willis at 15~$\mu$m and the ExaPipe benchmark use the PP set-up, and the right leg uses the VP set-up.

\subsection{Strong scaling}\label{sec:scaling}
\begin{figure}
\includegraphics[width=\textwidth]{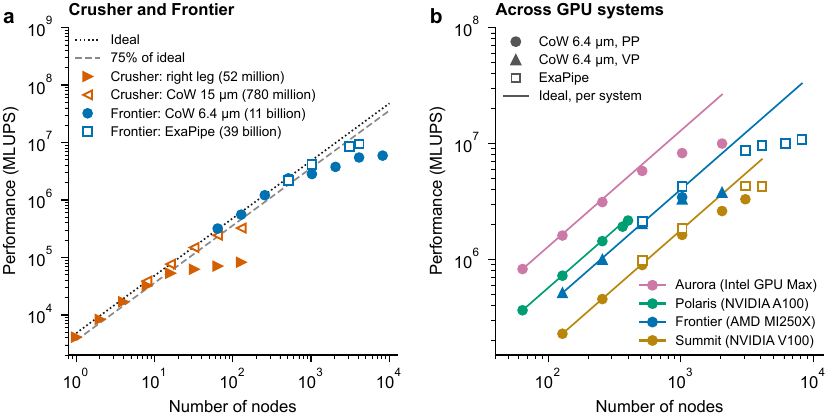}
\caption{\textbf{Parallel performance of HemeLB.} (\textbf{a}) Strong scaling on Crusher and Frontier for four problems of fixed size; the number of lattice sites is given in brackets. Dotted line, ideal linear scaling; dashed line, 75\% of ideal. (\textbf{b}) Strong scaling of the circle of Willis at 6.4~$\mu$m and the ExaPipe benchmark on Aurora, Polaris, Frontier and Summit. Filled circles, circle of Willis with the PP set-up; filled triangles, circle of Willis with the VP set-up; open squares, ExaPipe. Lines, ideal scaling from the smallest circle-of-Willis run on each system. Performance in MLUPS on logarithmic axes. The legend abbreviates the circle of Willis as CoW.}
\label{figure:scaling}
\end{figure}

A full-human computation is practical only if HemeLB uses a large machine efficiently. \Cref{figure:scaling} first reports strong scaling on Crusher and Frontier for four problems. Their sizes span nearly three orders of magnitude, from 52 million to 39 billion lattice sites.

The two largest problems test readiness for exascale hardware. For the circle of Willis at 6.4~$\mu$m, with 11 billion sites, performance follows the ideal line from 64 to 512 nodes. It continues to increase up to 8192 nodes, close to the full size of Frontier, and reaches about 6~TLUPS. Beyond 512 nodes, however, the gain from each added node decreases, and the points fall below the 75\% line from 1024 nodes onwards. The ExaPipe benchmark, with 39 billion sites, stays close to ideal scaling at 512 and 1024 nodes. It reaches about 10~TLUPS at 4096 nodes.

Smaller problems leave ideal scaling at smaller node counts. On Crusher, the circle of Willis at 15~$\mu$m, with 780 million sites, scales close to ideal from 8 to 64 nodes and falls below the 75\% line at 128 nodes. The right leg, with 52 million sites, follows the ideal line up to about 8 nodes. It then flattens and gains little beyond 32 nodes. Across all four series, scaling degrades once each GPU holds too few lattice sites. This pattern is consistent with the work per GPU becoming too small to hide communication costs~\cite{zacharoudiou2023development}.

\Cref{figure:scaling}b compares Aurora, Polaris, Frontier and Summit on the circle of Willis at 6.4~$\mu$m. At the smallest node counts, Aurora delivers about 13,000~MLUPS per node. The per-node rates are about 5,700~MLUPS on Polaris, 4,000~MLUPS on Frontier and 1,800~MLUPS on Summit. All four systems keep close to ideal scaling up to a few hundred nodes. The departure from that line then differs from machine to machine. Aurora falls to about 60\% of ideal at 1024 nodes. At 2048 nodes, Aurora reaches about 10~TLUPS, or 10 million MLUPS. Frontier keeps close to ideal scaling up to 512 nodes. At 1024 nodes it falls to about 80\% of ideal. Summit has the lowest rate per node and stays closest to the ideal line. It remains at about 90\% of ideal at 1024 nodes and at about 70\% at 2048 nodes. On Frontier, the PP and VP set-ups differ by only a few per cent at 1024 nodes. On the larger ExaPipe benchmark, Frontier reaches about 11~TLUPS at 8192 nodes. 

Two conclusions follow from \cref{figure:scaling}. First, HemeLB runs across the GPU architectures of current leadership systems. The same solver runs on AMD, Intel and NVIDIA GPUs, on the Frontier and Aurora exascale systems as well as on the earlier Summit and Polaris systems, and its scaling behaviour is the same on all of them. A HemeLB simulation is therefore not tied to one GPU vendor or one computing facility. Second, HemeLB takes LBM into the era of TLUPS. The LBM community has conventionally measured performance in MLUPS, and a single node in \cref{figure:scaling} delivers between about 1,800 and 13,000~MLUPS. On the largest runs, HemeLB sustains about 10~TLUPS, and it does so on two architectures: about 11~TLUPS for ExaPipe on 8192 Frontier nodes and about 10~TLUPS for the sparse circle of Willis on 2048 Aurora nodes. At this scale, LBM throughput is counted naturally in TLUPS rather than MLUPS. Strong scaling still ends once each GPU holds too few lattice sites, so on any of these machines the time to solution is set by the size of the problem.

\begin{table}[!htbp]
\def\baselinestretch{1}\fontsize{10}{12.5}\selectfont
\caption{\label{table:cost}Estimated cost of the full-human simulation with DNS at 20~$\mu$m and LES at 100~$\mu$m. The time step scales with the lattice spacing.}
\begin{ruledtabular}
\begin{tabular}{lrrr}
Quantity & DNS & LES & Factor \\
\hline
Lattice spacing, $\Delta x$ & 20~$\mu$m & 100~$\mu$m & 5 \\
Time step, $\Delta t$ & 1~$\mu$s & 5~$\mu$s & 5 \\
Fluid sites & $3.7\times10^{10}$ & $2.95\times10^{8}$ & 125 \\
Time steps per cardiac cycle & $10^{6}$ & $2\times10^{5}$ & 5 \\
Site updates per cardiac cycle & $3.7\times10^{16}$ & $5.9\times10^{13}$ & 625 \\
Node-hours per cardiac cycle & 7,700 & 12 & 625 \\
\end{tabular}
\end{ruledtabular}
\end{table}

\subsection{Cost of a full-human simulation}\label{sec:cost}
Problem size is the quantity that LES controls. For a fixed geometry, increasing the lattice spacing by factors of five and ten reduces the number of fluid sites by factors of about 125 and 1000. The time step changes with the lattice spacing as well. Numerical stability and accuracy require a small Mach number, which is proportional to $U\Delta t/\Delta x$ for a flow velocity $U$, so $\Delta t$ must scale with $\Delta x$. A 20~$\mu$m lattice therefore needs five times as many time steps per simulated second as a 100~$\mu$m lattice. For the full-human arterial tree (\cref{sec:fullhuman-setup}), the 100~$\mu$m lattice has 295 million fluid sites, whereas the 20~$\mu$m DNS lattice has 37 billion. Combining the two factors, a DNS costs about $125\times5=625$ times as much as the LES for each simulated second. \Cref{sec:aorta-validation} tests whether LES retains accuracy at these spacings on stenotic aortic flow.

\Cref{table:cost} estimates the cost of one cardiac cycle of the full-human simulation from the ExaPipe benchmark, which has about as many lattice sites as the full-human DNS. At the highest performance throughput, about 11~TLUPS on 8192 Frontier nodes, one 1~s cardiac cycle of DNS would take about an hour on nearly the whole machine, or about 7,700 node-hours for one patient. This estimate is likely to be optimistic. A sparse arterial tree runs below the throughput of the dense ExaPipe geometry, as the circle of Willis at 6.4~$\mu$m shows with about 6~TLUPS on the same number of nodes. The same cardiac cycle with LES requires about 12 node-hours. The largest cases in \cref{figure:scaling} show that HemeLB can still use most of an exascale machine when a problem demands it. With LES, the same machine can instead run many full-human simulations, for different patients or as ensembles for uncertainty quantification. Uncertainty quantification makes computational predictions more reliable. An ensemble of simulations shows how uncertainty in the inputs and parameters propagates to the predicted quantities, so that each prediction comes with an estimate of its own uncertainty~\cite{coveney2025molecular}, as in earlier HemeLB studies of aortic flow~\cite{lo2024uncertainty,xue2026uncertainty}.

\section{Validation on stenotic aortic flow}\label{sec:aorta-validation}
\begin{figure}
\includegraphics[width=\textwidth]{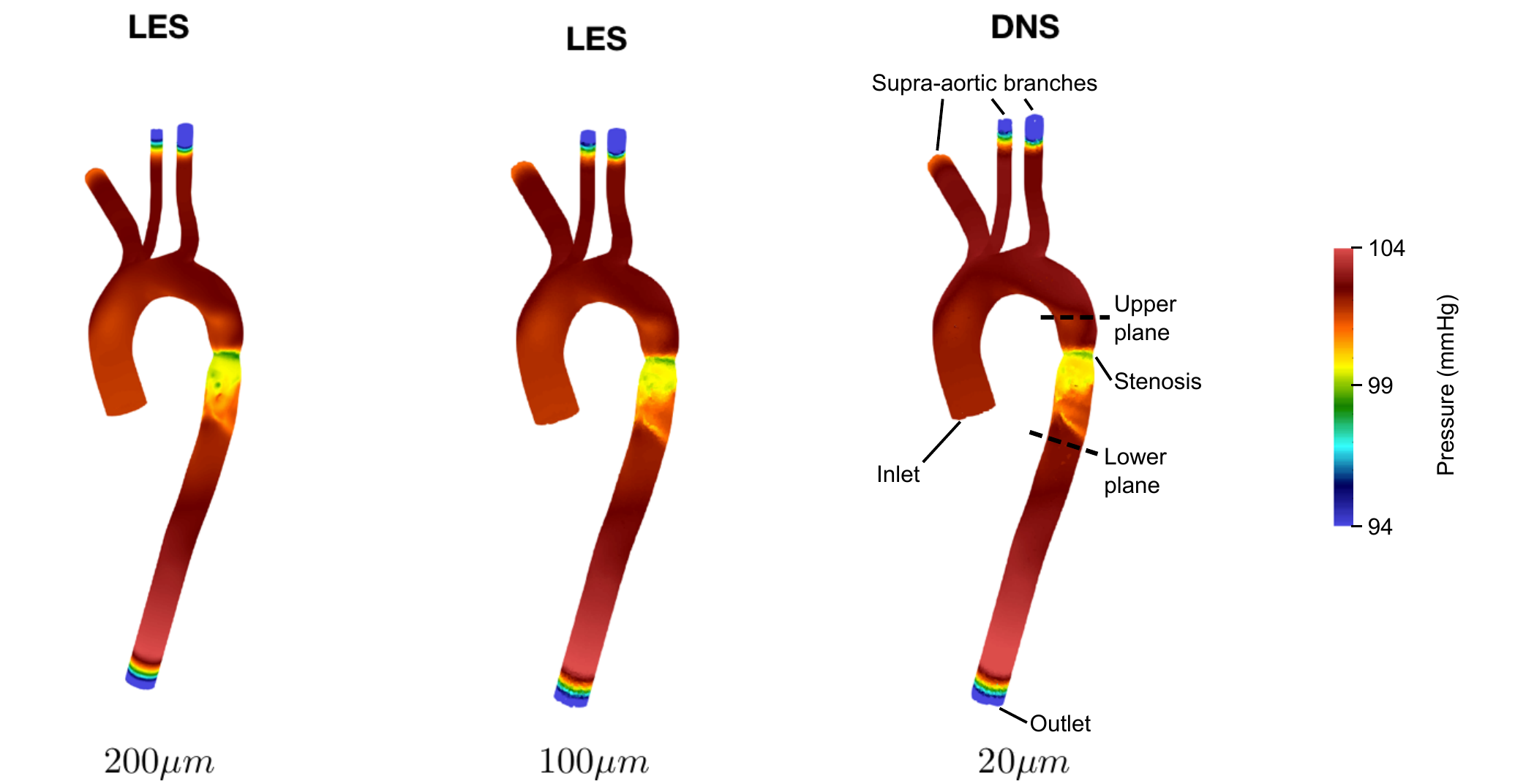}
\caption{\textbf{Stenotic aorta and wall pressure at three lattice resolutions.} Instantaneous pressure on the aortic wall for, from left to right, LES at 200~$\mu$m, LES at 100~$\mu$m and DNS at 20~$\mu$m. On the DNS geometry, labels mark the inlet, the supra-aortic branches, the outlet and the stenosis, and dashed lines mark the upper and lower planes on which pressure and velocity are compared. Colour scale, pressure in mmHg.}
\label{figure:coa_compare}
\end{figure}

\Cref{sec:cost} shows that computational cost is set by problem size, which a coarser lattice reduces. That reduction is useful only if the resolved hemodynamics are retained. Stenotic aortic flow supplies that test. The stenosis produces strong streamwise changes in pressure and velocity, so agreement with DNS on this geometry shows that the Smagorinsky closure has not sacrificed the large-scale solution. The DNS reference uses a spacing of 20~$\mu$m, and the two LES use 100~$\mu$m and 200~$\mu$m, five and ten times coarser. Instantaneous pressure on the aortic wall is compared first. Pressure and velocity magnitude on cross-sectional planes proximal and distal to the stenosis, marked in \cref{figure:coa_compare}, are then compared at 0.16~s within a cardiac cycle. Plane-averaged pressure on the same planes is then compared over one cardiac cycle, together with the error of each LES against the DNS.

\Cref{figure:coa_compare} compares the instantaneous pressure on the aortic wall. The geometry comprises the ascending aorta, the aortic arch with three supra-aortic branches, and the descending aorta, with the stenosis just distal to the arch. Blood enters at the ascending aorta and leaves through the three branches and the descending-aorta outlet, which are labelled on the DNS geometry together with the stenosis. In the DNS, pressure is high over the ascending aorta, the arch and the supra-aortic branches. It falls sharply across the stenosis to about 98--100~mmHg and recovers along the descending aorta. Both LES reproduce this distribution. The low-pressure region at the stenosis has the same location and a similar extent at 100~$\mu$m and 200~$\mu$m as at 20~$\mu$m. The main visible difference is a slightly higher pressure over the ascending aorta and arch in the DNS. The same small offset reappears in the plane-averaged pressure below.

\begin{figure}
\includegraphics[width=\textwidth]{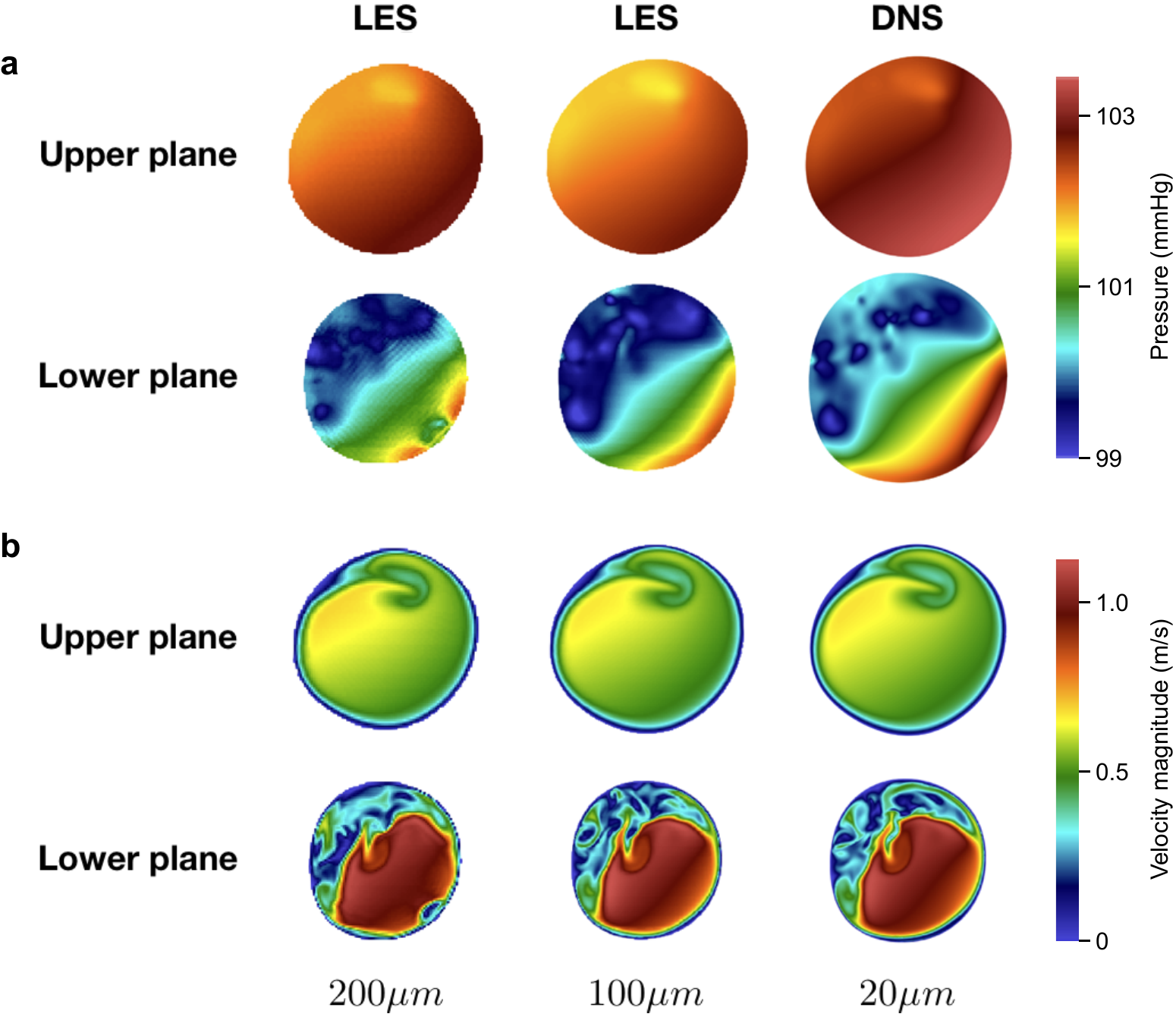}
\caption{\textbf{Cross-sectional pressure and velocity proximal and distal to the stenosis.} (\textbf{a}) Instantaneous pressure at 0.16~s on the upper plane, proximal to the stenosis, and on the lower plane, distal to the stenosis. (\textbf{b}) Velocity magnitude on the same planes at the same instant. Columns, from left to right: LES at 200~$\mu$m, LES at 100~$\mu$m and DNS at 20~$\mu$m.}
\label{figure:planes}
\end{figure}

\Cref{figure:planes} shows pressure and velocity magnitude at 0.16~s on two cross-sections of the descending aorta, which are marked by dashed lines in \cref{figure:coa_compare}. The upper plane lies just proximal to the stenosis, and the lower plane lies distal to it. On the upper plane, pressure is smooth and nearly uniform across the section at all three resolutions. The DNS gives slightly higher values than both LES, consistent with the wall pressure in \cref{figure:coa_compare}. The velocity on this plane has a moderate core of about 0.5--0.7~m\,s$^{-1}$ and a thin low-velocity layer along the wall. A curled shear layer near the top of the section has the same shape and position at 20, 100 and 200~$\mu$m. Proximal to the stenosis, both LES are therefore nearly indistinguishable from the DNS.

The lower plane is a more demanding test, because it cuts through the accelerated flow distal to the stenosis. In the DNS, this high-velocity region reaches a velocity magnitude of about 1~m\,s$^{-1}$ and occupies the larger part of the section. A narrow region of slower flow extends into it from its upper edge. A thin shear layer separates the high-velocity region from a low-velocity region that contains fine filamentary structures. The pressure field mirrors this structure. Pressure rises diagonally across the section, from about 99--100~mmHg in the low-velocity region to about 102.5--103~mmHg along the opposite wall. Small, isolated low-pressure spots within the low-velocity region are consistent with vortex cores.

Both LES reproduce the position and extent of the high-velocity region, the slower flow at its upper edge, the shear layer and the diagonal pressure gradient. Differences appear mainly at the smallest scales. The 100~$\mu$m LES retains most of the filaments seen in the DNS, whereas the 200~$\mu$m LES resolves fewer of them. Both LES resolutions give a broader low-pressure region than the DNS. At 200~$\mu$m, the high-pressure band along the wall is also weaker, and lattice texture is visible on the plane. These differences concern structures that the subgrid model represents rather than resolves. They do not change the position of the high-velocity region or the direction of the cross-sectional pressure gradient.

\begin{figure}
\includegraphics[width=\textwidth]{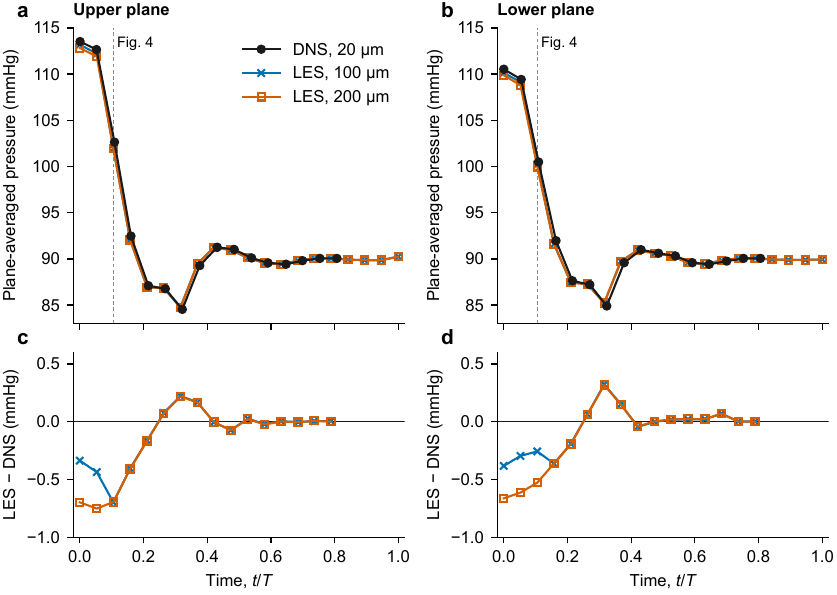}
\caption{\textbf{Plane-averaged pressure over one cardiac cycle and LES error against DNS.} (\textbf{a}) Pressure averaged over the upper plane, proximal to the stenosis, against time $t$ normalised by the cardiac period $T$. (\textbf{b}) The same for the lower plane, distal to the stenosis. The dashed lines mark the instant of the cross-sectional comparison, 0.16~s. (\textbf{c},\textbf{d}) Difference between each LES and the DNS at the same output step on the planes of \textbf{a} and \textbf{b}. The DNS record ends at $t/T\approx0.8$.}
\label{figure:pressure_cycle}
\end{figure}

\Cref{figure:pressure_cycle} follows the pressure averaged over each plane through one cardiac cycle. On both planes, pressure falls from its maximum at the start of the cycle to a minimum of about 85~mmHg near $t/T=0.32$, where $T$ is the cardiac period. The maximum is about 113~mmHg on the upper plane and about 110~mmHg on the lower plane. Pressure then rises briefly to about 91~mmHg near $t/T=0.43$ and settles close to 90~mmHg for the rest of the cycle. The three resolutions follow this waveform closely, and their curves nearly overlap throughout. The dashed line marks the instant shown in \cref{figure:planes}, which lies on the steep decline that follows the peak.

The lower panels of \cref{figure:pressure_cycle} show the error against the DNS. It is largest at the pressure peak, where both LES lie below the DNS. This offset matches the higher proximal pressure of the DNS in \cref{figure:coa_compare,figure:planes}. The largest difference is 0.70~mmHg for the 100~$\mu$m LES and 0.75~mmHg for the 200~$\mu$m LES, both on the upper plane. This is below 1\% of the local pressure and about 2.5\% of the pressure variation over the cycle. On the lower plane, the peak error of the 100~$\mu$m LES is smaller than that of the 200~$\mu$m LES, 0.38 against 0.66~mmHg. After the decline, from $t/T\approx0.2$, the two LES differ from the DNS by at most about 0.3~mmHg, near the pressure minimum, and by less than 0.1~mmHg for $t/T>0.4$.

\Crefrange{figure:coa_compare}{figure:pressure_cycle} show that coarsening the lattice by factors of five and ten retains the large-scale aortic pressure and velocity. The remaining differences concern the smallest vortical structures distal to the stenosis and a pressure offset of at most 0.75~mmHg near the peak. This result supports using the same LES formulation for the full-human computation.

\section{Full-human large-eddy simulation of the arterial system}\label{sec:fullhuman}
The validated LES (\cref{sec:aorta-validation}) and the parallel performance of HemeLB (\cref{sec:scaling}) together make a three-dimensional computation of the whole human arterial tree possible. This section describes the geometry and numerical set-up of the full-human model, the hemodynamics over the whole body and in each body region, and the limitations of the present model.
\begin{figure}
\includegraphics[width=\textwidth]{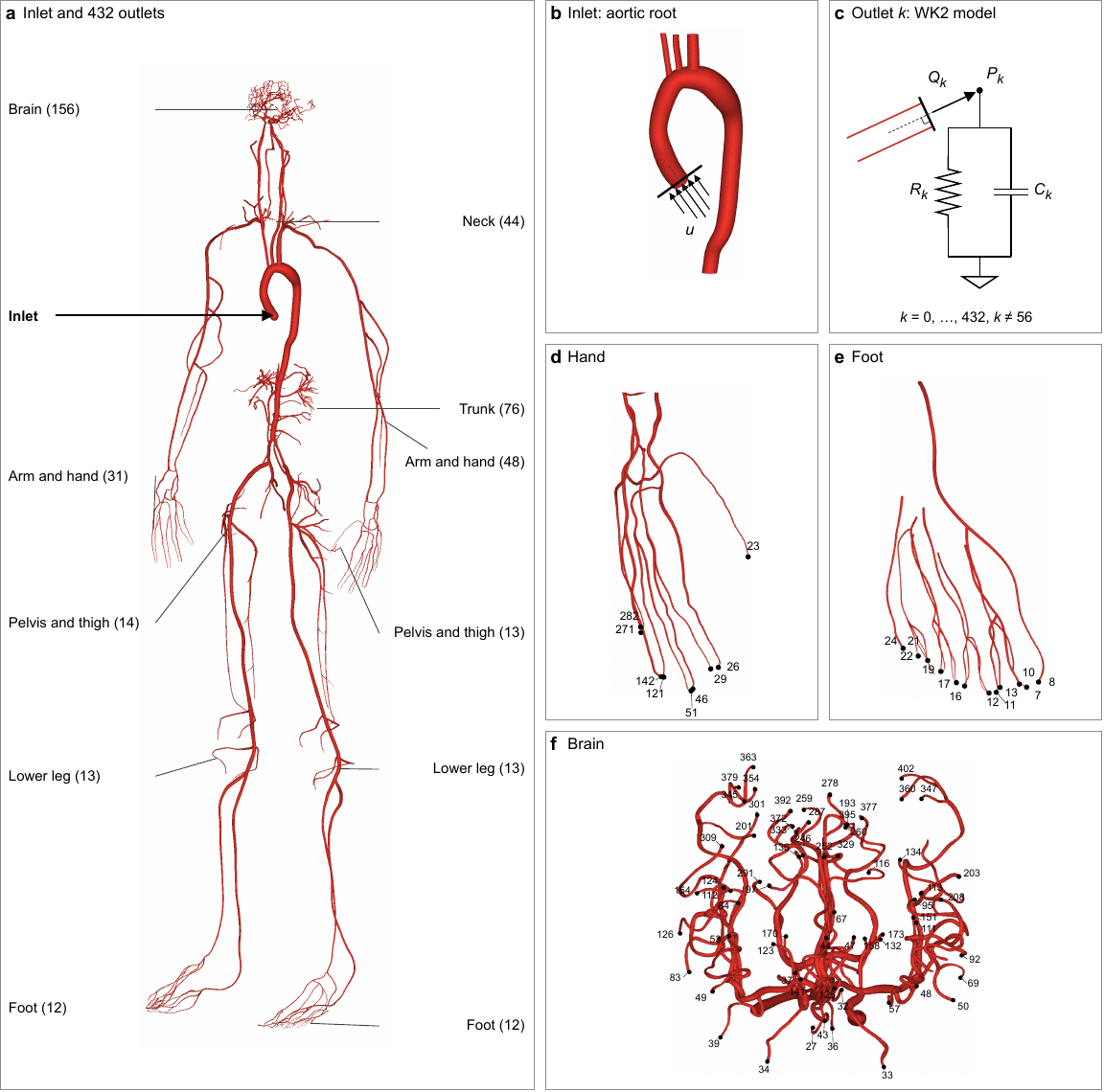}
\caption{\textbf{Numerical set-up of the full-human model.} (\textbf{a}) Whole arterial tree, turned by 45$^\circ$ from the front view. The arrow marks the single inlet at the aortic root, and the labels give the number of the $N_{\mathrm{out}}=432$ outlets in each body region. (\textbf{b}) Inlet at the aortic root. The vessel is cut by a plane perpendicular to its axis, and a pulsatile velocity with a parabolic profile is imposed on the cut. (\textbf{c}) Outlet $k$. The openings are numbered from 0 to 432 as in the simulation; opening 56 is the inlet, and $k$ runs over the other 432 numbers. The vessel is likewise cut perpendicular to its axis, and the outflow $Q_k$ through the cut sets the outlet pressure $P_k$ through a WK2 model with resistance $R_k$ and compliance $C_k$. (\textbf{d}--\textbf{f}) Hand, forefoot and cerebral arteries with representative outlets numbered.}
\label{figure:fullhuman_setup}
\end{figure}

\subsection{Geometry and numerical setup}\label{sec:fullhuman-setup}
\Cref{figure:fullhuman_setup} summarises the set-up of the full-human model. \Cref{figure:fullhuman_setup}a shows the whole arterial tree, which extends from the head and neck through the upper limbs and the thoracic and abdominal aorta to the lower limbs and feet. The aorta is the trunk from which all regions branch. The carotid and vertebral arteries of the neck, which arise from the aortic arch and the subclavian arteries, lead to the cerebral arteries. The subclavian arteries continue through the arms to the arteries of the hands, which extend into the fingers. Along the abdominal aorta, the renal and visceral branches arise before the aorta divides into the iliac arteries. These supply the pelvis and continue through the legs to the arteries of the feet, which extend into the toes. The model therefore retains small distal arteries in every region. The geometry is taken from the extended cardiac-torso human phantom developed at Duke University~\cite{segars20104d}. Its anatomy was segmented from the anatomical images of the Visible Human Project and from patient computed tomography data. In the reconstructed arterial tree, the smallest vessels have a diameter of 400~$\mu$m. The aortic root is the only inlet, and every other vessel end is an outlet, $N_{\mathrm{out}}=432$ in total. The labels in \cref{figure:fullhuman_setup}a give the number of outlets in each body region.

The inlet and the outlets are made by cutting the vessels of the surface model in 3D Slicer~\cite{fedorov20123d}, as demonstrated in \cref{figure:fullhuman_setup}b,c. The end point of each vessel is first marked on the surface model. At each end point, the local direction of the vessel is estimated from its centreline, and the vessel is cut by a plane perpendicular to this direction. The cut face is then a cross-section of the vessel with its normal along the vessel axis, so the inlet profile is imposed across the vessel and the outlet flow rate is the flow along it. An oblique cut would instead produce an elongated face whose normal deviates from the axis. The cut surface model is then voxelised on the lattice of HemeLB. \Cref{figure:fullhuman_setup}b shows the inlet at the aortic root, where a pulsatile velocity waveform with a parabolic profile is imposed. The waveform is the heartbeat profile that McCullough and Coveney imposed at the inlets of the circle of Willis~\cite{mccullough2023high}, which was obtained from Patronis et al.\ at a heart rate of 68 beats per minute~\cite{patronis2018modeling}. Here it is rescaled to a centreline peak of $U_{\mathrm{peak}}=1$~m\,s$^{-1}$ and to a cardiac period of 1~s. At each outlet, sketched in \cref{figure:fullhuman_setup}c, the cut sets only the position and the area $A_k$, and the share of the flow that leaves through the outlet is controlled by its WK2 model, defined by \cref{eq:wk2} in \cref{sec:method-wk}. The resistance $R_k$ and compliance $C_k$ are set so that the peak inflow of the imposed profile, $Q_{\mathrm{peak}}=\tfrac{1}{2}U_{\mathrm{peak}}A_{\mathrm{in}}$ with $A_{\mathrm{in}}$ the area of the inlet, divided among the outlets in proportion to their areas, produces the same pressure, $P_{\mathrm{t}}=20$~mmHg, at every outlet, and all outlets share the time constant $\tau_{\mathrm{wk}}=0.05$~s:
\begin{equation}
\label{eq:wk2_parameters}
R_k=\frac{P_{\mathrm{t}}}{Q_{\mathrm{peak}}}\,\frac{\sum_{l} A_l}{A_k},\qquad C_k=\frac{\tau_{\mathrm{wk}}}{R_k},
\end{equation}
where the sum runs over all $N_{\mathrm{out}}$ outlets. The simulation starts from rest with $P=0$ at every outlet. \Cref{figure:fullhuman_setup}d--f show that the model ends in the small arteries of the fingers, the toes and the brain, and label representative outlets with their opening indices in the simulation.

\begin{figure}[h]
\includegraphics[width=\textwidth]{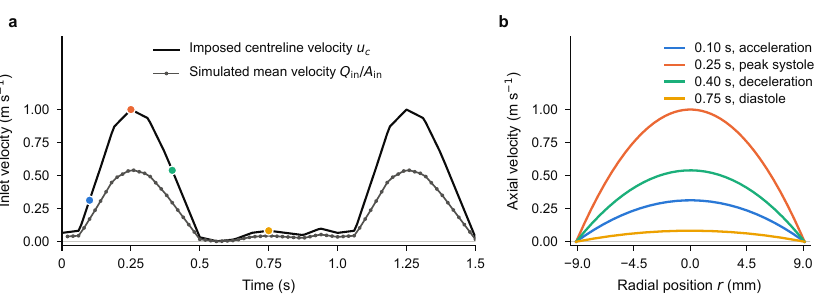}
\caption{\textbf{Inlet boundary condition at the aortic root.} (\textbf{a}) Imposed centreline velocity $u_c$ and simulated mean velocity over the inlet during the first 1.5~s of the simulation, which starts from rest. Black line, imposed; grey dots, simulated, one per output step of 0.02~s. The imposed waveform repeats with a period of 1~s, and coloured dots mark the instants shown in \textbf{b}. (\textbf{b}) Imposed parabolic profile across the inlet, with cap radius $R_{\mathrm{in}}=9.11$~mm, at $t=0.10$~s in acceleration, 0.25~s at peak systole, 0.40~s in deceleration and 0.75~s in diastole.}
\label{figure:inlet}
\end{figure}

\Cref{figure:inlet}a shows two velocities at the inlet. One is the imposed centreline velocity, $u_c(t)$, which peaks at 1~m\,s$^{-1}$. The other is the simulated mean velocity over the inlet, $Q_{\mathrm{in}}/A_{\mathrm{in}}$, where $Q_{\mathrm{in}}$ is the inflow rate; it peaks at 0.54~m\,s$^{-1}$, about half the centreline peak, as expected for a parabolic profile. \Cref{figure:inlet}b shows this parabolic profile across the inlet at four instants of the cycle.

% \XX{[List $R_k$ and $C_k$ for every outlet in a supplementary table.]}

The smallest vessels in the full-human model, 400~$\mu$m in diameter, have a radius of 200~$\mu$m. This radius spans two lattice spacings at 100~$\mu$m but only one at 200~$\mu$m, and a single lattice spacing cannot resolve a physically meaningful velocity profile across the vessel. The 100~$\mu$m resolution is also more accurate: in the validation of \cref{sec:aorta-validation}, the 100~$\mu$m LES is closer to the DNS than the 200~$\mu$m LES, both in the vortical structures distal to the stenosis and in the plane-averaged pressure. The full-human computation therefore uses the 100~$\mu$m resolution. The 100~$\mu$m lattice has about eight times as many fluid sites as the 200~$\mu$m lattice, but still about 125 times fewer than a 20~$\mu$m lattice. At 100~$\mu$m, the full-human arterial tree has 295 million fluid sites. This simulation uses a time step of $\Delta t=4$~$\mu$s, so one cardiac cycle of 1~s takes $2.5\times10^5$ time steps. Each cardiac cycle costs about 15--18 node-hours with 2 nodes from Frontier. The simulation covers 1.5~s from rest, $3.75\times10^5$ time steps, which is enough for the flow to become periodic.
\begin{figure}
\includegraphics[width=\textwidth]{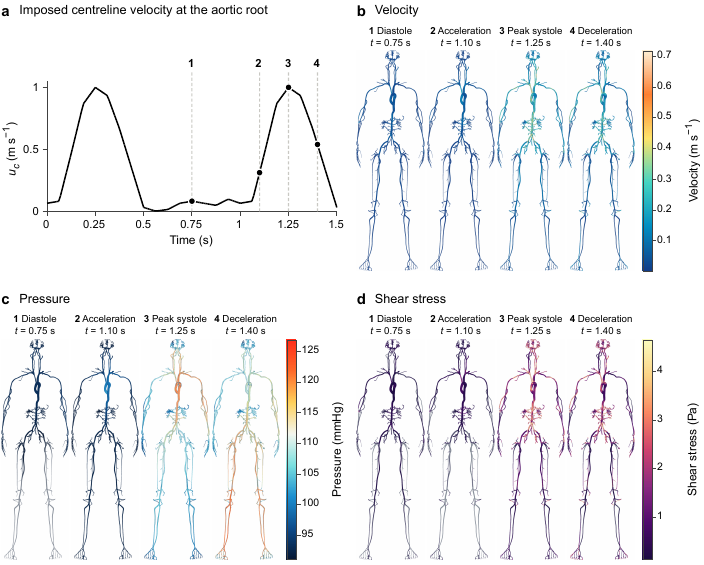}
\caption{\textbf{Velocity, pressure and shear stress over the full-human arterial tree at four instants of the cardiac cycle.} Computed with LES at $\Delta x=100$~$\mu$m. (\textbf{a}) Imposed centreline velocity at the aortic root; dots 1--4 mark the instants shown in \textbf{b}--\textbf{d}: diastole, $t=0.75$~s; acceleration, $t=1.10$~s; peak systole, $t=1.25$~s; and deceleration, $t=1.40$~s. (\textbf{b}) Velocity magnitude $|\boldsymbol{u}|$. (\textbf{c}) Pressure with a reference value of 90~mmHg. (\textbf{d}) Near-wall viscous shear stress magnitude. Each field uses the same colour scale at all four instants.}
\label{figure:fields_peak}
\end{figure}
\subsection{Whole-body hemodynamics}\label{sec:fullhuman-whole}
\Cref{figure:fields_peak} shows the velocity, the pressure and the shear stress over the whole arterial tree at four instants, marked on the inlet waveform in \cref{figure:fields_peak}a: diastole at $t=0.75$~s, and the acceleration, peak and deceleration of the second systole at $t=1.10$, 1.25 and 1.40~s. To quantify the four snapshots, we group the lattice sites of each snapshot by body region from their position: the aorta, the head and neck, the upper limbs, the visceral organs and the lower limbs. We further divide the lower limbs by height above the soles into the thighs, the lower legs and the feet. The values below are medians over the sites of a region unless stated otherwise.

All three fields follow the inlet pulse from the heart to the periphery. \Cref{figure:fields_peak}b shows that the velocity is low throughout the tree in diastole, with a median of 0.06~m\,s$^{-1}$ in the aorta and 0.01--0.02~m\,s$^{-1}$ in the other regions, and that during acceleration it rises first in the aorta and in the head and neck, whereas the lower limbs remain almost at rest. At peak systole, the median velocity is 0.44~m\,s$^{-1}$ in the aorta, 0.21~m\,s$^{-1}$ in the head and neck and 0.13--0.16~m\,s$^{-1}$ in the upper limbs, the viscera and the thighs, but only 0.05~m\,s$^{-1}$ in the lower legs and 0.02~m\,s$^{-1}$ in the feet. During deceleration, the aortic velocity falls to 0.29~m\,s$^{-1}$, whereas the 95th percentile of the velocity in the feet rises from 0.08 to 0.21~m\,s$^{-1}$ as the systolic flow arrives there. The pressure pulse in \cref{figure:fields_peak}c travels the same way and grows as it does so. At peak systole, the pressure falls from 126~mmHg in the aorta to 100~mmHg in the feet. During deceleration, 0.15~s later, the gradient reverses: the pressure falls to 114~mmHg in the aorta but rises in the lower legs and the feet, which turn from blue at peak systole to orange and become the highest-pressure region of the tree. \Cref{figure:fields_peak}d shows that the shear stress concentrates in the branches rather than in the aorta. At peak systole, the visceral branches and the head and neck reach the top of the colour range, whereas the 95th percentile in the aorta, which reflects the near-wall values, is 2.6~Pa, where the wide lumen keeps the velocity gradient at the wall small. During deceleration, it rises in the feet from 1.1 to 3.0~Pa as the systolic flow arrives.

\subsection{Regional hemodynamics}\label{sec:fullhuman-regional}
\begin{figure}
\includegraphics[width=\textwidth]{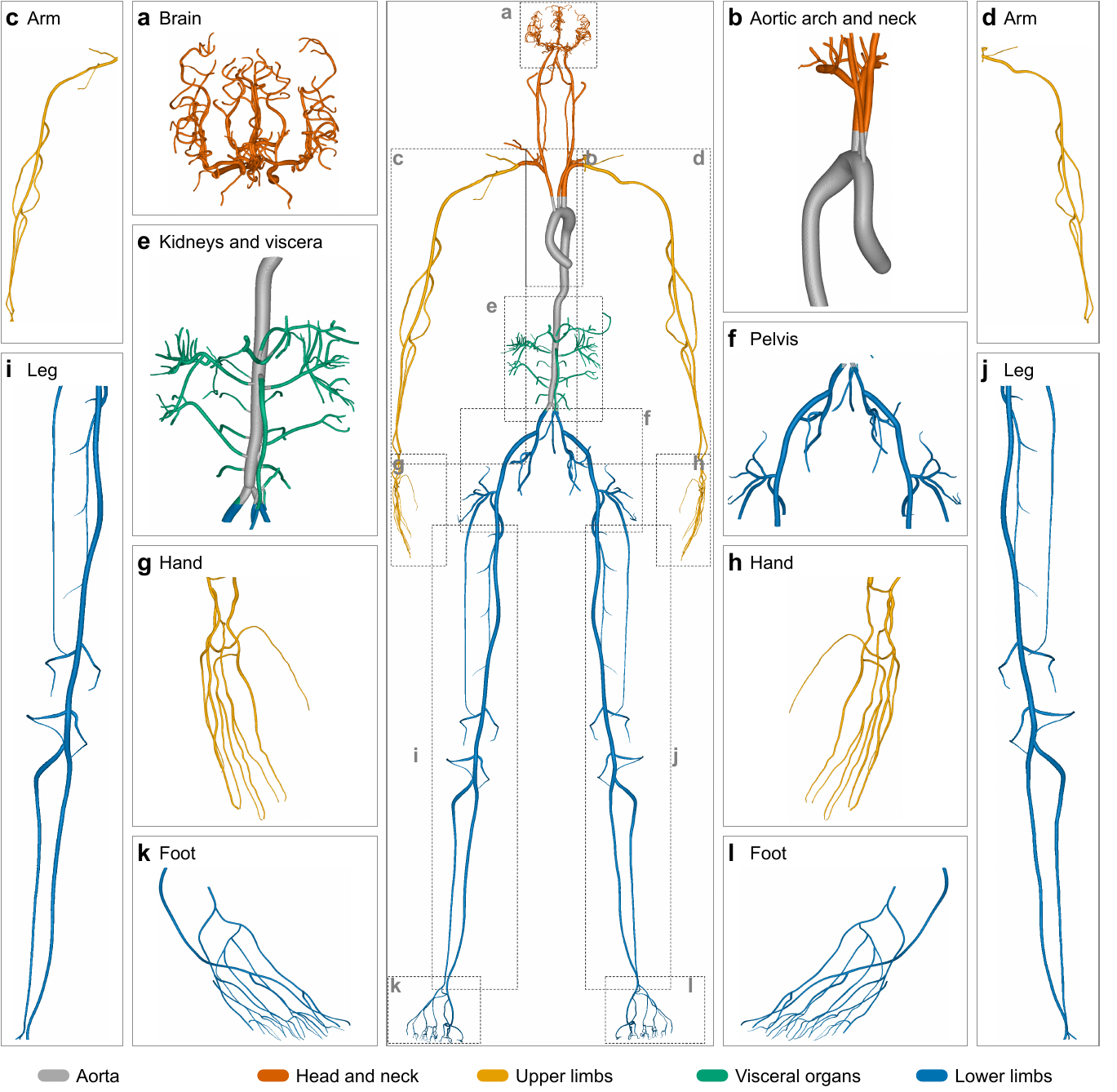}
\caption{\textbf{Arterial geometry of the full-human model.} Centre, front view of the whole arterial tree, coloured by body region: aorta, head and neck, upper limbs, visceral organs and lower limbs. Boxes bounded by dotted lines mark the close-up views. (\textbf{a}) Cerebral arteries, viewed from the front and above. (\textbf{b}) Aortic arch and neck arteries, viewed obliquely. (\textbf{c},\textbf{d}) Arteries of the arms. (\textbf{e}) Abdominal aorta with the renal and visceral branches. (\textbf{f}) Aortic bifurcation and pelvic arteries; (\textbf{g},\textbf{h}) Hands; (\textbf{i},\textbf{j}) Legs; (\textbf{k},\textbf{l}) Feet.}
\label{figure:fullhuman_geometry}
\end{figure}
This section examines the local hemodynamics of each body region of the full-human arterial network. The outlets are grouped by the region they perfuse: the head and neck, the upper limbs, the visceral organs supplied by the abdominal aorta, and the lower limbs. \Cref{figure:fullhuman_geometry} shows the arterial tree coloured by these regions, with the aorta in grey as the trunk from which they all branch, and close-up views of the vessels in each region. \Cref{figure:fullhuman_geometry}a shows the cerebral arteries, which are supplied by the neck arteries that arise from the aortic arch in \cref{figure:fullhuman_geometry}b. \Cref{figure:fullhuman_geometry}c,d show the arteries of the two arms, \cref{figure:fullhuman_geometry}e the abdominal aorta with its renal and visceral branches, and \cref{figure:fullhuman_geometry}f the aortic bifurcation, where the iliac arteries arise and supply the pelvis. The remaining close-ups show the distal ends of the limbs: the hands in \cref{figure:fullhuman_geometry}g,h, the legs in \cref{figure:fullhuman_geometry}i,j and the feet in \cref{figure:fullhuman_geometry}k,l. 
\begin{figure}
\includegraphics[width=\textwidth]{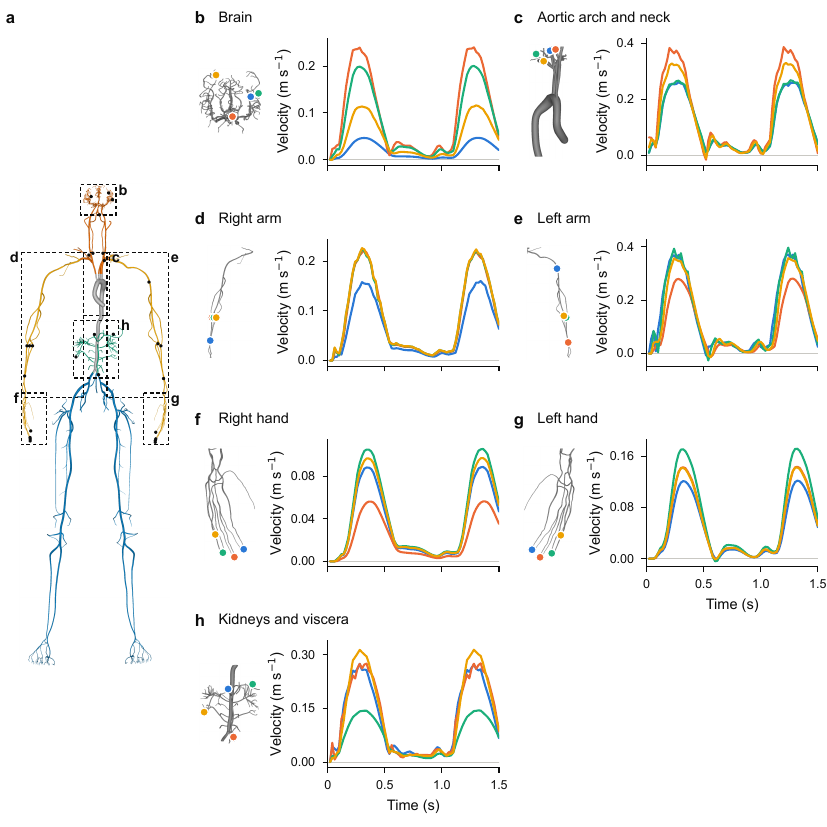}
\caption{\textbf{Outlet velocity in the upper body.} (\textbf{a}) Front view of the arterial tree, coloured by body region: aorta, head and neck, upper limbs, visceral organs and lower limbs. Boxes bounded by the dashed lines mark the regions of panels \textbf{b}--\textbf{h}, and coloured dots mark the representative outlets. The right side of the body is on the left side of the view. (\textbf{b}--\textbf{h}) Mean velocity, the outflow divided by the cap area, at four outlets per region over the first 1.5~s from rest. Each close-up marks the outlets in the colours of their curves.}
\label{figure:velocity_upper}
\end{figure}

To compare the regions, we follow the mean velocity at four representative outlets in each region over the first 1.5~s of the simulation, which starts from rest. The flow becomes periodic within the first cycle, so the quantities below are evaluated over the last full period, $0.5\le t<1.5$~s, and times are given from the start of each cycle. Each cycle has two phases. In systole, during the first half of the cycle, the inlet velocity rises steeply to its peak and then decelerates; in diastole, during the second half, only a small forward flow remains. The outlets follow the same phases, but with region-specific delays and waveform shapes.

In the upper body, \cref{figure:velocity_upper} shows that the velocity peaks are highest at the outlets nearest the heart, 0.26--0.39~m\,s$^{-1}$ in the neck branches of the aortic arch, and lowest at the most distal outlets, 0.06--0.17~m\,s$^{-1}$ in the hands. In the brain, the viscera, the arms and the hands, the velocity remains essentially positive over the whole cycle, and its diastolic value is a small fraction of the systolic peak, as at the inlet.

\begin{figure}
  \includegraphics[width=\textwidth]{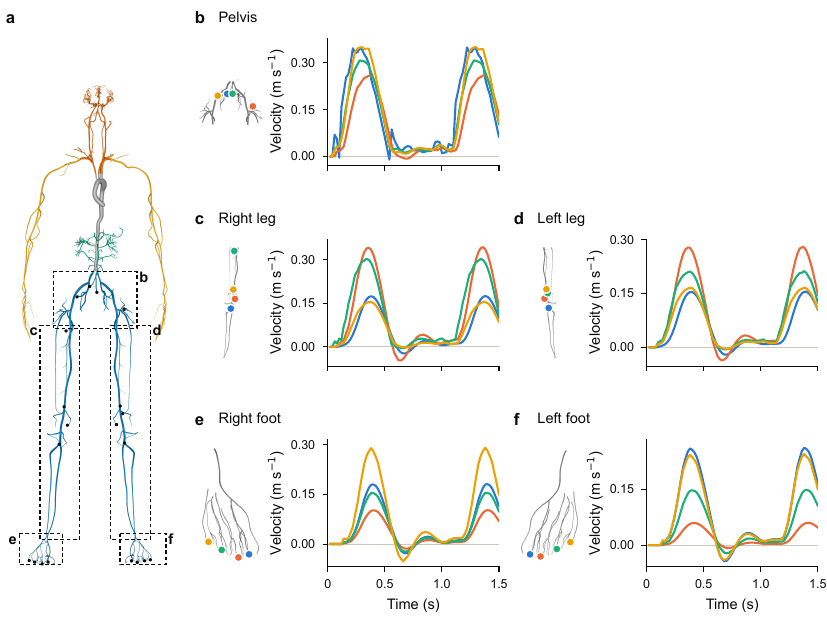}
  \caption{\textbf{Outlet velocity in the lower body.} (\textbf{a}) Front view of the arterial tree, coloured by body region. Boxed bounded by dotted lines mark the regions of panels \textbf{b}--\textbf{f}, and coloured dots mark the representative outlets. The right side of the body is on the left side of the view. (\textbf{b}--\textbf{f}) Mean velocity, the outflow divided by the cap area, at four outlets per region over the first 1.5~s from rest. Each close-up marks the outlets in the colours of their curves. Negative values mark reverse flow into the model.}
  \label{figure:velocity_lower}
\end{figure}

In the lower body, \cref{figure:velocity_lower} shows a different waveform: every outlet of the legs and feet has a triphasic waveform. Forward flow in systole is followed by reverse flow in early diastole, at $t\approx0.64$--$0.72$~s, which lasts 0.08--0.18~s and reaches $-0.05$~m\,s$^{-1}$, up to 18\% of the systolic peak in the feet. At most outlets, a second, weaker forward wave then peaks at $t\approx0.84$--$0.92$~s. The pulsatility, the range of the velocity over the period relative to its mean, also rises towards the feet. These observations agree with Doppler measurements in the arteries of the human leg, which show a triphasic waveform~\cite{kim2020interpretation} and a pulsatility that increases distally~\cite{hendrickx1994grundlagen}. The two legs have similar peaks, 0.15--0.34~m\,s$^{-1}$ on the right and 0.15--0.28~m\,s$^{-1}$ on the left.

\begin{figure}
\includegraphics[width=\textwidth]{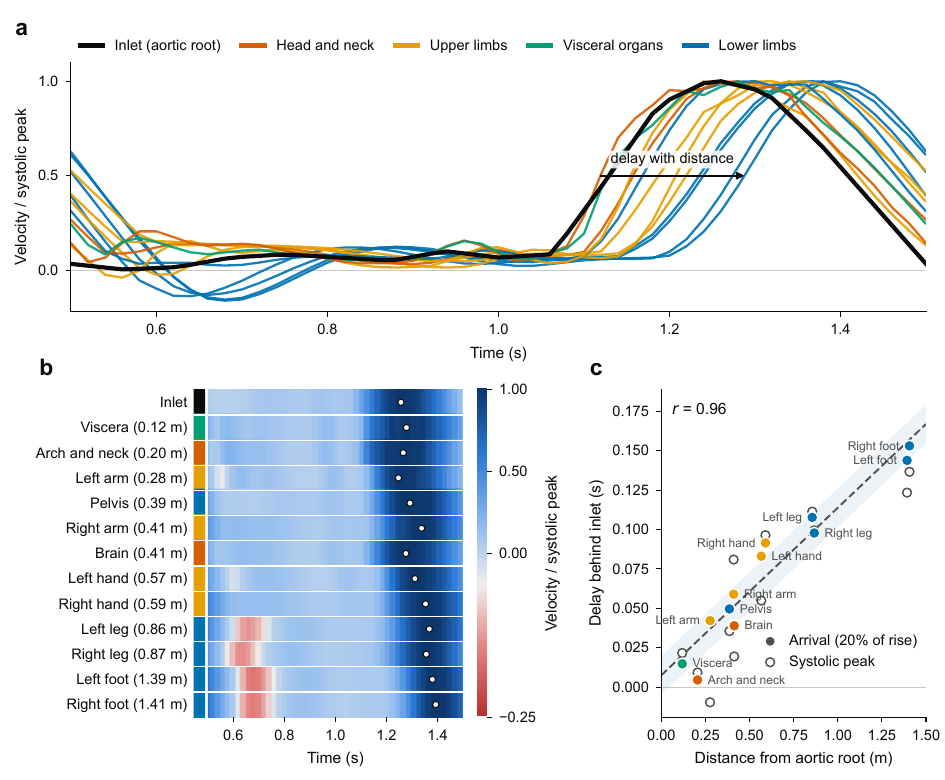}
\caption{\textbf{Phase lag of the outlet velocity behind the inlet.} Each body region is represented by one outlet with a large share of its outflow. (\textbf{a}) Mean velocity divided by its systolic peak over the last period, $0.5\le t<1.5$~s, at the inlet and at the 12 outlets. Black, inlet; colours, body region of each outlet, as in \cref{figure:fullhuman_geometry}. The arrow marks the delay of the systolic rise from the neck to the feet. (\textbf{b}) The same as a heat map, with rows ordered by the straight-line distance of each outlet from the aortic root, given in brackets; the strip on the left gives the body region of each row, red marks reverse flow, and white dots mark the systolic peaks. (\textbf{c}) Delay behind the inlet against distance. The arrival of the pulse is the instant at which the velocity completes 20\% of its systolic rise. Filled symbols: arrival of the pulse, coloured by body region; open symbols: systolic peak; dashed line and shaded band: linear fit to the arrival delays and the standard deviation of its residuals.}
\label{figure:phase_lag}
\end{figure}

The pulse reaches the outlets later with increasing distance from the heart. \Cref{figure:phase_lag} summarises this phase lag with one representative outlet per region, chosen among those with a large share of the regional outflow. We take the arrival time as the instant at which the velocity has completed 20\% of its systolic rise. \Cref{figure:phase_lag}a,b shows that, normalised by their own systolic peaks, the waveforms keep their shape but move to later times with distance from the heart. As \cref{figure:phase_lag}c shows, the arrival delay grows nearly linearly with the straight-line distance from the aortic root, with a correlation coefficient of 0.96, and reaches 0.15~s in the feet. The systolic peak follows with a similar delay. Because the walls are rigid, this delay arises from the inertia of the blood in the long distal vessels, the compliance of the WK2 models and the weak compressibility of the LBM fluid.

These results lead to two conclusions. First, the regional differences, namely the later arrival of the pulse with distance and the triphasic flow in the lower limbs, arise from the three-dimensional arterial geometry alone, because every outlet uses the same WK2 design. A full-human model therefore captures interactions between regions that are lost when each region is simulated separately with prescribed boundary conditions. Second, a single simulation resolves the hemodynamics of every region simultaneously, from the cerebral arteries to the arteries of the feet. Regions as distant as the brain and the feet can therefore be compared at the same instant and under the same inflow, which separate regional models, each with its own boundary conditions, cannot provide.

\subsection{Limitations and future work}\label{sec:limitations}

The present model has several limitations, which also indicate directions for future work. First, the vessel walls in the full-human computation are rigid, so the model cannot represent an elastic pulse wave. The arrival times and distal pressure amplification in \cref{sec:fullhuman-whole,sec:fullhuman-regional} arise from blood inertia, the compliance of the WK2 outlets and the weak compressibility of the LBM fluid. Elastic wall boundary conditions have not yet been applied to the full-human computation~\cite{mccullough2021efficient}. Future work will apply them, with regionally varying wall stiffness, so that elastic pulse-wave propagation can be studied. Second, the effects of ageing are not considered. The model represents a single arterial geometry, whereas the arterial walls stiffen with age and the aorta becomes more tortuous~\cite{xiao2013multi}. A study of ageing would therefore require both age-dependent wall properties and geometries from subjects of different ages. Third, we set the Windkessel parameters using a generic distribution of flow across the outlets, without calibration to patient data. They do not account for the resistance of the arterial paths or the low resistance of the cerebral vascular bed, so the regional distribution of flow is not expected to be physiological. Their time constant, $\tau_{\mathrm{wk}}=0.05$~s, is also short compared with the cardiac period of 1~s, so the WK2 models store little volume, and the diastolic flow at the outlets follows the inlet waveform rather than being sustained by peripheral compliance. Region-specific resistances and compliances, calibrated against measured flow fractions and waveforms~\cite{lo2022parametric}, are therefore needed before the regional hemodynamics can be compared with clinical data. Finally, validation is a crucial step for any model. The LES has so far been validated only against DNS in the stenotic aorta, and the full-human results have not yet been compared with clinical measurements. In future work, flow waveforms at sites along the arterial tree will be compared with Doppler ultrasound and four-dimensional flow magnetic resonance imaging~\cite{kamada2022blood}, a step that is needed before the approach can inform clinical practice.

% \subsection{Whole human CVD impact due to the stenotic of the aorta}

\section{Conclusion}\label{sec:conclusion}
This work has followed a path from computational capability to a practical whole-body simulation. We first showed that HemeLB runs with the same scaling behaviour on AMD, Intel and NVIDIA GPUs, including the Frontier and Aurora exascale supercomputers, and that it sustains about 10~TLUPS on both. The LBM community has conventionally measured performance in MLUPS. With this throughput, HemeLB takes LBM into the era of TLUPS. This capability alone does not make full-human simulation useful. A lattice fine enough to resolve the aortic flow directly would make each run too expensive for the many simulations that patient-specific studies require. We therefore combined HemeLB with Smagorinsky LES, whose locality preserves the scalability of LBM on GPUs.

In stenotic aortic flow, LES on lattices five and ten times coarser than the 20~$\mu$m DNS reproduced the wall pressure, the cross-sectional pressure and velocity proximal and distal to the stenosis, and the plane-averaged pressure over the cardiac cycle. The largest difference in plane-averaged pressure was below 1\% of the local pressure, and the remaining differences were confined to the smallest vortical structures distal to the stenosis. At the same time, the two LES reduced the computational cost relative to the DNS by factors of about 600 and 4800. Coarsening the lattice therefore did not sacrifice the accuracy of the large-scale hemodynamics.

This reduction makes a full-human computation affordable. We computed blood flow from the head to the feet, including the arteries of the brain, the viscera, the hands and the feet, in a model with one inlet at the aortic root and 432 Windkessel outlets, at a lattice spacing of 100~$\mu$m set by the narrowest retained vessels. One cardiac cycle cost about 15--18 node-hours on two Frontier nodes, and the flow became periodic within the first cycle. Over the whole tree, the pressure pulse grew as it travelled from the aorta to the feet, and the shear stress concentrated in the branches rather than in the aorta. At the outlets, the waveforms reproduced regional features that arise from the arterial geometry alone: the pulse arrived later with distance from the heart, by about 0.15~s in the feet, and the flow in the legs and feet was triphasic, as in Doppler measurements of the human leg. Because a single simulation resolves every region simultaneously, regions as distant as the brain and the feet can be compared at the same instant and under the same inflow. The lower cost per simulation brings ensembles of whole-body simulations within reach, as required for patient-specific studies with uncertainty quantification~\cite{coveney2025molecular,lo2024uncertainty}.

Beyond addressing the limitations discussed above, the framework can be extended in several directions. Coupling the arterial tree to the venous circulation would close the systemic loop. A whole-body model also allows the systemic influence of a local CVD, such as a stenosis of the aorta, to be studied, together with diseases that involve the vascular network as a whole, for example the effect of peripheral arterial disease on flow in an abdominal aortic aneurysm~\cite{lo2024uncertainty}. In the same way, future studies could examine how an ischaemic stroke, such as an occlusion of the basilar artery in the circle of Willis~\cite{mccullough2023high}, alters cerebral blood flow and how this change affects flow in the rest of the body.

\section{Acknowledgements}

We acknowledge funding support from European Commission CompBioMed Centre of Excellence (Grant No. 675451 and 823712). Support from the UK Engineering and Physical Sciences Research Council under the projects ``UK Consortium on Mesoscale Engineering Sciences (UKCOMES)" (Grant No.
EP/R029598/1) and ``Software Environment for Actionable and VVUQ-evaluated Exascale Applications (SEAVEA)" (Grant No. EP/W007711/1) is gratefully acknowledged. 
This research used resources of the Oak Ridge Leadership Computing Facility at the Oak Ridge National Laboratory, which is supported by the Office of Science of the U.S. Department of Energy under Contract No. DE-AC05-00OR22725. We thank Jon W. S. McCullough, Sharp C. Y. Lo and Bálint Joó for useful discussions. We also thank Ivan Benemerito and Alberto Marzo of the Insigneo Institute at the University of Sheffield for valuable discussions during their visit to University College London in 2025.

\appendix
\section{Notation and abbreviations}\label{app:notation}
\Cref{table:notation} lists the symbols used in this paper, grouped by the model in which they are introduced, together with their meanings and units. \Cref{table:abbreviations} lists the abbreviations.

\begin{table}[!htbp]
\def\baselinestretch{1}\fontsize{10}{12.5}\selectfont
\caption{\label{table:notation}Symbols used in this paper, grouped by the model in which they are introduced, with their meanings and units. LU, lattice units. A dash marks a dimensionless quantity or an index.}
\begin{ruledtabular}
\begin{tabular}{lll}
Symbol & Description & Unit \\
\hline
\multicolumn{3}{l}{\textit{Lattice Boltzmann method (\cref{sec:method-lbm})}} \\
$\mathbf{x}$, $t$ & Position of a lattice site and time & m, s \\
$\Delta x$ & Lattice spacing & $\mu$m \\
$\Delta t$ & Lattice Boltzmann time step & s \\
$\mathbf{c}_i$ & Discrete velocity in direction $i\in\{0,\ldots,18\}$ & LU \\
$f_i$, $f_i^{eq}$ & Distribution function and its equilibrium & LU \\
$w_i$ & Weight of direction $i$ & -- \\
$\Omega$ & BGK collision operator, $\Delta t/\tau$ & -- \\
$\tau$ & Relaxation time & LU \\
$c_s$ & Lattice speed of sound & LU \\
$\rho$ & Fluid density & LU \\
$\mathbf{u}$ & Fluid velocity & m\,s$^{-1}$ \\
$\nu$ & Kinematic viscosity set by $\tau$ (\cref{eq:nu}) & m$^2$\,s$^{-1}$ \\
\multicolumn{3}{l}{\textit{Smagorinsky subgrid-scale model (\cref{sec:method-sgs})}} \\
$\nu_0$ & Molecular kinematic viscosity of blood & m$^2$\,s$^{-1}$ \\
$\nu_t$ & Turbulent (subgrid-scale) viscosity & m$^2$\,s$^{-1}$ \\
$\nu_{\mathrm{eff}}$ & Effective viscosity, $\nu_0+\nu_t$ & m$^2$\,s$^{-1}$ \\
$C_{\mathrm{smag}}$ & Smagorinsky constant & -- \\
$\Delta$ & Filter width & m \\
$|\bar{\mathbf{S}}|$ & Magnitude of the filtered strain-rate tensor & s$^{-1}$ \\
\multicolumn{3}{l}{\textit{Windkessel outlets (\cref{sec:method-wk})}} \\
$P$ & Gauge pressure at an outlet & mmHg \\
$Q$ & Outward volumetric flow rate at an outlet & mL\,s$^{-1}$ \\
$R$ & Peripheral resistance of a WK2 model & kg\,m$^{-4}$\,s$^{-1}$ \\
$C$ & Compliance of a WK2 model & m$^4$\,s$^2$\,kg$^{-1}$ \\
$n$ & Time-step index & -- \\
$\rho_{\mathrm{out}}$ & Outlet lattice density, $1+P/c_s^2$ & LU \\
\multicolumn{3}{l}{\textit{Sponge layer (\cref{sec:sponge-sgs})}} \\
$\nu_s$ & Kinematic viscosity in the sponge zone & m$^2$\,s$^{-1}$ \\
$p$ & Viscosity ratio at the outlet & -- \\
$\|d\|$ & Distance to the outlet plane & m \\
$w_{\mathrm{width}}$ & Width of the sponge zone & m \\
\multicolumn{3}{l}{\textit{Validation (\cref{sec:aorta-validation})}} \\
$T$ & Cardiac period & s \\
\multicolumn{3}{l}{\textit{Full-human set-up (\cref{sec:fullhuman-setup})}} \\
$N_{\mathrm{out}}$ & Number of outlets in the full-human model & -- \\
$k$, $l$ & Outlet indices & -- \\
$R_k$, $C_k$ & Resistance and compliance of outlet $k$ & as $R$, $C$ \\
$A_k$ & Area of outlet $k$ & m$^2$ \\
$A_{\mathrm{in}}$ & Area of the aortic root & m$^2$ \\
$U_{\mathrm{peak}}$ & Centreline peak inlet velocity & m\,s$^{-1}$ \\
$u_c$ & Imposed centreline inlet velocity & m\,s$^{-1}$ \\
$R_{\mathrm{in}}$ & Radius of the inlet cap & mm \\
$Q_{\mathrm{in}}$ & Volumetric inflow rate at the aortic root & mL\,s$^{-1}$ \\
$Q_{\mathrm{peak}}$ & Peak inflow, $\tfrac{1}{2}U_{\mathrm{peak}}A_{\mathrm{in}}$ & mL\,s$^{-1}$ \\
$P_{\mathrm{t}}$ & Target outlet pressure at peak inflow & mmHg \\
$\tau_{\mathrm{wk}}$ & Windkessel time constant, $R_kC_k$ & s \\
\end{tabular}
\end{ruledtabular}
\end{table}

\begin{table}[!htbp]
\def\baselinestretch{1}\fontsize{10}{12.5}\selectfont
\caption{\label{table:abbreviations}Abbreviations used in this paper.}
\begin{ruledtabular}
\begin{tabular}{ll}
Abbreviation & Meaning \\
\hline
BGK & Bhatnagar--Gross--Krook (collision operator) \\
CFD & Computational fluid dynamics \\
CoW & Circle of Willis \\
CPU & Central processing unit \\
CVD & Cardiovascular disease \\
D3Q19 & Three-dimensional lattice with 19 discrete velocities \\
DNS & Direct numerical simulation \\
GPU & Graphics processing unit \\
LBM & Lattice Boltzmann method \\
LES & Large-eddy simulation \\
LU & Lattice units \\
MLUPS & Million lattice-site updates per second \\
PP & Pressure--pressure (boundary-condition set-up) \\
SGS & Subgrid scale \\
TLUPS & Trillion lattice-site updates per second \\
VP & Velocity--pressure (boundary-condition set-up) \\
WK2 & Two-element Windkessel model \\
\end{tabular}
\end{ruledtabular}
\end{table}
\clearpage

%\bibliographystyle{splncs04}
\bibliography{ref_Mendeley}

\end{document}